\documentclass[preprint,showpacs,amsmath,amssymb]{revtex4}

\usepackage{graphicx}
\usepackage{dcolumn}
\usepackage{bm}
\usepackage{setspace}

\begin{document}

\title{Discovery of Bragg confined hybrid modes with high Q-factor in a hollow dielectric resonator}
\author{Jean-Michel le Floch, Michael E. Tobar, David Mouneyrac}
\affiliation{\\ University of Western Australia School of Physics M013, 35 Stirling Hwy. Crawley 6009 WA, Australia}
\email{mike@physics.uwa.edu.au}
\author{Dominique Cros}
\affiliation{\\ XLIM - UMR CNRS no. 6172 - FacultŽ des Sciences et Techniques - 123, avenue A. Thomas, 87060 Limoges Cedex, France\\}
\author{Jerzy Krupka}
\affiliation{\\ Institute of Microelectronics and Optoelectronics, Warsaw University of Technology, Koszykowa  75, Warsaw, Poland}

\date{\today}

\begin{abstract}
The authors report on observation of Bragg confined mode in a hollow cylindrical dielectric cavity. A resonance was observed at $13.4$ $GHz$ with an unloaded Q-factor of order $2\times10^5$, which is more than a factor of 6 above the dielectric loss limit. Previously such modes have only been realized from pure Transverse Electric modes with no azimuthal variations and only the $E_{\phi}$ component. From rigorous numeric simulations it is shown that the mode is a hybrid mode with non-zero azimuthal variations and with dominant $E_r$ and $E_{\phi}$ electric field components and $H_z$ magnetic field component.
\end{abstract}

\pacs{41.20.-q, 77.22.-d, 42.60.Da}
\keywords{Bragg reflection, Dielectric resonator, high Q-factor, microwaves}
\maketitle

Low noise oscillators require high-Q resonators for low phase noise and high stability. The Q-factor of a standard dielectric resonator is usually limited by the dielectric loss-tangent of the material. One way of beating the loss-tangent limit is by building a layered structure that confines the mode mainly in the central air region through Bragg reflection. In the past this has solely been achieved with $TE_{m,n,p}$ modes of azimuthal mode number $m = 0$ in cylindrical resonators\cite{Maggiore,Comte1,Comte2,Flory97,Flory98,Tobar05}  ($n$ is the radial mode number and $p$ the axial). This is because the mode has only an $E_{\phi}$ electric field that remains tangential to both the radial and axial surfaces, which is a requirement of Bragg reflection. In this work we show the discovery of modes of $m > 0$, which also maintain Bragg confinement in hollow cylindrical structures. The modes are hybrid (all field components), however, the dominant fields for these modes are $E_r$, $E_{\phi}$ and $H_z$, and Bragg reflection is allowed as the radial component of the Electric field exists near the central region of the resonator and supplies a tangential boundary condition to the axial Bragg reflectors. In contrast the azimuthal electric field exists mainly at all boundaries of the Bragg reflectors and Bragg confinement of the mode is also achieved in the radial direction.

\begin{figure}
\centering
\includegraphics[width=3.3in]{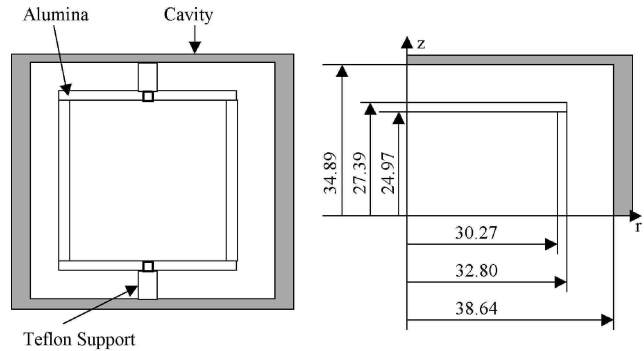} 
\caption{Left: Schematic of the cavity showing the hollow cylindrical Alumina structure supported by Teflon posts loaded inside a silver plated copper cavity. Right: Top right quadrant of the structure, showing dimensions.}
\label{fig:1}
\end{figure}

A hollow single-layer alumina cylinder of height $49.94\ mm$ and diameter $65.6\ mm$ was manufactured and supported by Teflon spaces in a cylindrical metallic cavity (see Fig.\ref{fig:1}). In this work, the $m = 1$ hybrid mode was measured to have a frequency of $13.4 GHz$ with a Q-factor of order $2\times10^5$, which is more than $6$ times the dielectric loss limit. In contrast single layered cylindrical structures using the lowest order fundamental $m = 0$ modes have only achieved about a factor of $2$\cite{Tobar05,Tobar04,Krupka05}. The field structure and properties were verified using Method of Lines. This result presented here is also greater than a high-Q sapphire Whispering Gallery mode resonator at room temperature\cite{wgmode}, but achieved using a cheaper material with the potential to enable the constructing equivalent state-of-the-art low noise oscillators\cite{HartnettEL,Woode,Tobar94}.

\begin{figure}
\centering
\includegraphics[width=3.3in]{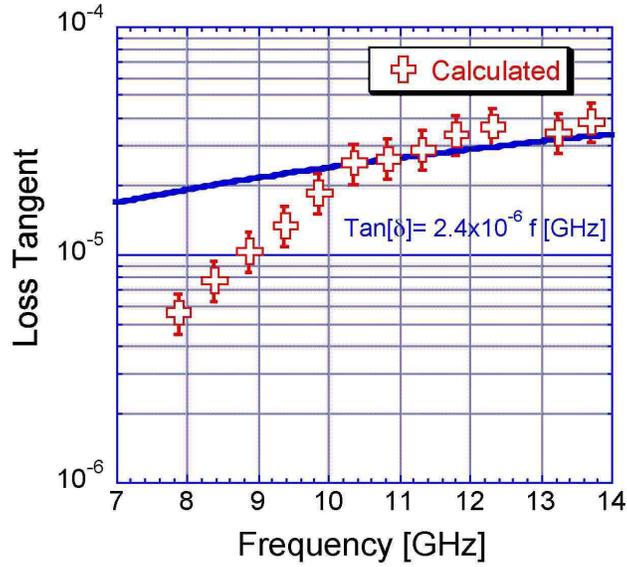} 
\caption{Calculated and fitted loss tangent from the results presented in Tab.\ref{Tab1}, and the calculation from Eq.\ref{eq1}. The calculation is only accurate above 10 GHz when $Q^{-1} > R_s/G$. Typically in this regime unloaded Q-factor measurements are of order $20\%$ accurate, which is reflected in the error bars.}
\label{fig:2}
\end{figure}

To characterize the dielectric properties of the alumina we use the whispering gallery mode method\cite{Tobar98,Krupka99,Krupka99b}. Modes are simulated using Method of Lines software to predict the frequency, geometric factor and filling factor of the fundamental $WGH$ mode family. Simulations are compared with measurements in order to estimate the loss tangent and the permittivity of the alumina sample. Results are given in Tab.\ref{Tab1} and Fig.\ref{fig:2}. To calculate the loss tangent as a function of frequency, the results in Tab.\ref{Tab1} are combined with Eq.\ref{eq1} and plotted in Fig.\ref{fig:2} (note the filling factor in Teflon is very small and can be ignored). 

\begin{equation}
tan[\delta]=\frac{Q^{-1}_{meas}-\frac{R_s}{G}}{p_e}
\label{eq1}
\end{equation}
Hear $R_s$ is the cavity surface resistance, which was determined to be   from the Q-factor of the   mode in the empty cavity. The calculation in Eq.\ref{eq1} is only accurate when $Q^{-1} > R_s/G$, which is the case in Fig.\ref{fig:2} when $f < 10 GHz$. Above $10 GHz$ the usual frequency dependence of the loss tangent is measured, and determined to be $2.4\times10^{-6} f$.

\begin{figure}
\centering
\includegraphics[width=3in]{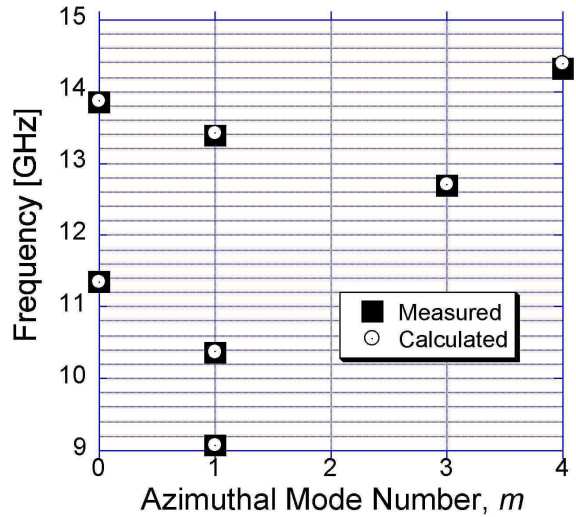} 
\caption{Calculated and measured frequencies of some higher order confined Bragg modes.}
\label{fig:3}
\end{figure}

\begin{figure}
\centering
\includegraphics[width=3in]{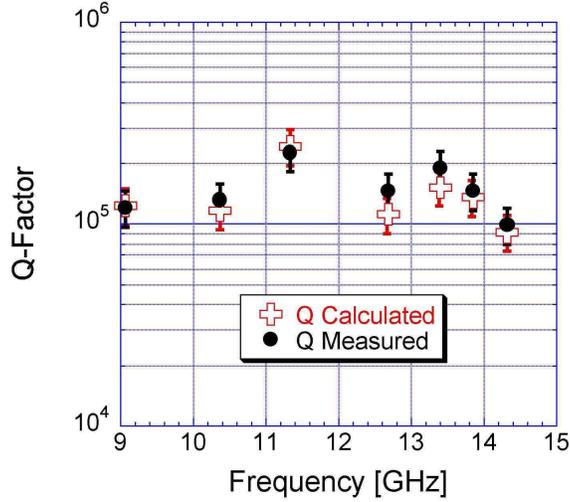} 
\caption{Measured and calculated unloaded Q-factors of the Bragg modes in  Fig.\ref{fig:3}. The measurement is made in by measuring the bandwidth in transmission in the low coupling regime, see  Fig.\ref{fig:6} , for example. Typically this types of measurement are about 20\% accurate.}
\label{fig:4}
\end{figure}

\begin{figure}
\centering
\includegraphics[width=3.3in]{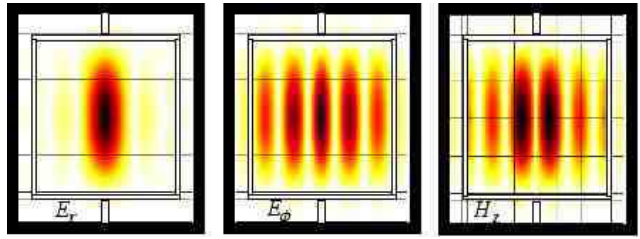} 
\caption{Density plot (modulus squared) of the dominant electric and magnetic field components for the $13.4 GHz$ mode of $m = 1$ as calculated using the Method of Lines (see Tab.\ref{Tab1} for filling factors). The hollow alumina cylinder is outlined (along with the Teflon supports), which confines the field in the internal free space region}
\label{fig:5}
\end{figure}

\begin{figure} 
\centering
\includegraphics[width=3.3in]{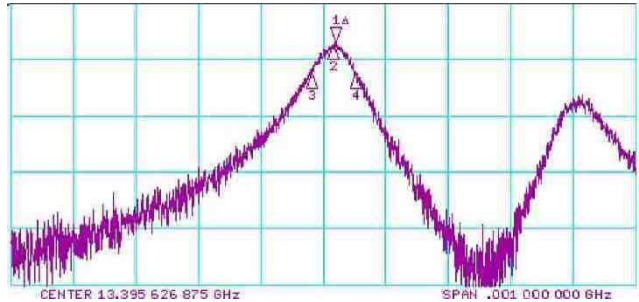} 
\caption{Measurement of the Q-factor of the $m =1$ Bragg mode. Resonance frequency is $13.39564 GHz$, loss on resonance is high ($-68.4 dB$) illustrating low coupling, and measured resonance bandwidth is $69.836 kHz$ giving a unloaded Q-factor of $192,000$. Note the mode exists as a doublet due to the degeneracy for even (cosine) and odd (sine) mode dependence along the azimuth when $m > 0$.}
\label{fig:6}
\end{figure}

Several confined higher order Bragg modes were measured in the structure with Q-factors of order $10^5$ or larger. Modes were identified after the characterization of the permittivity and loss-tangent and results are shown in Fig.\ref{fig:3} and \ref{fig:4}. In previous work only Bragg confined modes of $m = 0$ have been characterized, and here we unequivocally identify Bragg confined modes of azimuthal mode number $m > 0$. The highest Q-factors are measured for the $11.34\ GHz$ $(Q = 2.25\times10^5)$ and $13.40\ GHz$ $(Q = 1.91\times10^5) $  modes (see Fig.\ref{fig:5} for a plot of the field density and Fig.\ref{fig:6} for a network analyzer measurement of the Q-factor). Given that we measured the loss tangent of the alumina to be $tan\delta_{Alumina} = 2.4\times10^{-6} f[GHz]$ both modes are a factor of $6.1$ above the dielectric loss limit. We may also compare the results to a sapphire whispering gallery $WGH$ mode. Given that the loss tangent parallel to the c-axis is $tan\delta_{Sapphire}  = 4.2\times10^{-7} f^{1.09}[GHz]$\cite{Hartnett} the $11.34$ and $13.4 GHz$ modes have Q-factors of $1.3$ and $1.4$ times greater than sapphire $WGH$ modes respectively. 

The temperature coefficient of frequency for such resonators are also expected to be an order of magnitude lower than sapphire resonators, as they will be dominated by the temperature coefficient of expansion rather than the temperature coefficient of permittivity\cite{Krupka05}. Also, some of the modes show a doublet structure similar to Whispering Gallery modes due to the cosine-sine degeneracy for modes of $m > 0$. This may effect the noise performance of an oscillator\cite{Woode} due to competition between modes, but generally can be avoided with correct phase control within the loop of an resonator-oscillator\cite{HartnettEL}.

In summary, we have shown for the existence of Bragg modes with greater than zero azimuthal variations in a hollow ceramic Alumina structure. Higher order Bragg modes were measured with a Q-factor of $30$ to $40 \%$ greater to that of sapphire $WGH$ modes, and with a factor of $6.1$ above the dielectric loss limit of the alumina dielectric. 

\begin{acknowledgments}
This work was funded by the Australian Research Council.
\end{acknowledgments}

\newpage

\begin{table}
\caption{\label{Tab1} Characteristics of the $WGH_{m,0,0}$ mode family ($m$ is the azimuthal mode number), including measured $f_{meas}$, and calculated frequency $f_{calc}$ $[GHz]$, measured Q-factor $Q$, and calculated electric energy filling factor  $p_e$, and G-factor, $G$. The permittivity is estimated to be $9.73$ to allow agreement with the calculated and measured frequencies, the material loss tangent is calculated using Eq.\ref{eq1} and shown in Fig.\ref{fig:2}}
\begin{ruledtabular}
\begin{tabular}{ccccccc}
$m$ & $f_{meas}$ & $Q$ &$ f_{calc}$ & $p_{e}$ & $G$\\
\hline
$9$ & $7.881$ & $4.33\times10^4$ & $7.877$ &$0.880$ & $1.31\times10^3$\\
$10$ & $8.384$ & $4.81\times10^4$ & $8.376$ &$0.889$ & $1.76\times10^3$\\
$11$ & $8.877$ & $5.02\times10^4$ & $8.873$ &$0.898$ & $2.38\times10^3$\\
$12 $ & $9.379$ & $4.94\times10^4$ & $9.368$ &$0.906$ & $3.23\times10^3$\\
$13 $ & $9.860$ & $4.32\times10^4$ & $9.861$ &$0.913$ & $4.40\times10^3$\\
$14 $ & $10.357$ & $3.60\times10^4$ & $10.352$ &$0.919$ & $6.00\times10^3$\\
$ 15$ & $10.839$  & $3.60\times10^4$ & $10.840$ &$0.925$ & $8.20\times10^3$\\
$ 16$ & $11.328$  & $3.40\times10^4$ & $11.325$ &$0.931$ & $1.12\times10^4$\\
$ 17$ & $11.801$  & $3.00\times10^4$ & $11.808$ &$0.936$ & $1.54\times10^4$\\
$ 18$ & $12.320$  & $2.80\times10^4$ & $12.289$ &$0.940$ & $2.12\times10^4$\\
$ 19$ & $-$  & $-$ & $12.767$ &$0.945$ & $2.91\times10^4$\\
$ 20$ & $13.248$  & $3.00\times10^4$ & $13.244$ &$0.949$ & $4.01\times10^4$\\
$ 21$ & $13.715$  & $2.70\times10^4$ & $13.718$ &$0.952$ & $5.52\times10^4$\\
\end{tabular}
\end{ruledtabular}
\end{table}

\end{document}